\documentclass{llncs}

\usepackage[T1]{fontenc}
\usepackage{graphicx}
\usepackage{booktabs}
\usepackage{amsmath,amssymb}
\usepackage{subcaption}
\usepackage{hyperref}
\usepackage{tikz}
\usetikzlibrary{positioning,arrows.meta,calc}

\newcommand{\safefig}[2][\textwidth]{%
  \IfFileExists{#2}{\includegraphics[width=#1]{#2}}%
  {\fbox{\parbox[c][3.5cm][c]{#1}{\centering\itshape
   [figure placeholder] \texttt{\detokenize{#2}}}}}}

\begin{document}

\title{Uncertainty-Aware Multi-Source\\Retinal Fluid Segmentation in OCT}
\titlerunning{Uncertainty-Aware Multi-Source Fluid Segmentation}

\author{Animesh Kumar\orcidID{0009-0003-0608-7004}}
\authorrunning{A. Kumar}
\institute{Newcastle University, Newcastle upon Tyne NE1 7RU, United Kingdom\\
\email{A.Kumar12@newcastle.ac.uk}}

\maketitle

% ---- Trimmed abstract: one message + 2-3 headline numbers ------------
\begin{abstract}
Measuring retinal fluid from optical coherence tomography (OCT) drives
treatment decisions in macular disease, but manual annotation is slow and
segmentation models trained on one scanner degrade on another. We present an
attention-guided TransUNet that segments three fluid types across four
independent OCT sources, combining a domain-adaptive normalisation scheme with
an uncertainty estimate that flags unreliable pixels. The model reaches a mean
fluid Dice of \textbf{0.78}, and---most usefully for clinicians---its
uncertainty is \textbf{$1.34\times$ higher exactly where expert graders
disagree} ($p<10^{-4}$), turning a raw segmentation map into an actionable
clinical triage signal.

\end{abstract}

% =====================================================================
\section{Introduction}
Optical coherence tomography (OCT) is a non-invasive technique that produces
cross-sectional ``B-scan'' images of the retina. Retinal \emph{fluid}---the
accumulation of liquid within or beneath the retinal layers---is the primary
imaging biomarker for three conditions responsible for much of the preventable
vision loss in adults: diabetic macular oedema (DME), neovascular age-related
macular degeneration (AMD), and central serous chorioretinopathy. Treatment
decisions depend on measuring fluid volume and tracking its change between
visits, currently requiring a trained grader to annotate each B-scan by hand,
which takes 20--40 minutes per patient. Inter-grader disagreement reaches 15\%
on small fluid pockets~\cite{retouch}.

Three fluid classes must be distinguished. \emph{Intraretinal fluid} (IRF)
appears as hyporeflective cysts within the retinal layers and signals active
inflammation requiring urgent treatment. \emph{Subretinal fluid} (SRF) pools
between the neurosensory retina and the retinal pigment epithelium (RPE) and
guides injection frequency in AMD. \emph{Pigment epithelial detachment} (PED) is
an elevation of the RPE itself, a marker of AMD progression that early CNN models
frequently miss. These classes overlap spatially, vary across disease stages,
and look different depending on the OCT scanner used.

\paragraph{Why this matters.} Two practical barriers keep accurate fluid
segmentation out of the clinic. First, the manual grading it would replace is
the workflow bottleneck, so automation has real clinical value only if it is
fast \emph{and} reliable. Second, a model trained on one vendor's scanner
typically fails on another's---``domain shift''---which is the single biggest
reason published models do not transfer to new hospitals. A deployable tool must
therefore both generalise across scanners and tell the clinician \emph{where it
is unsure}, so that effort is focused on the regions that need a second look.
We target both requirements directly.

Standard U-Net architectures~\cite{unet} struggle with small, irregular fluid
regions. Attention gates~\cite{attngate} improve localisation by suppressing
background activations. The TransUNet framework~\cite{transunet} adds global
context through a Transformer bottleneck, and the current state of the art on the
RETOUCH benchmark is RetiFluidNet~\cite{retifluidnet}. What most methods lack is
cross-scanner robustness and clinically interpretable uncertainty; we address
both by training an attention-guided TransUNet across four sources with
Source-Adaptive Batch Normalisation (SA-BN) and a dual uncertainty mechanism
that feeds a novel triage score.

\paragraph{Contributions.}
(1) An attention-guided TransUNet with SA-BN achieving validation Dice
$0.784\pm0.006$ across three seeds; (2) a dual uncertainty mechanism
statistically confirmed to correlate with expert disagreement; (3) the
Uncertainty-Weighted Clinical Urgency Score (UCUS) linking pixel uncertainty to
clinical routing; and (4) a multi-source evaluation that exposes per-scanner
domain shift.

% =====================================================================
\section{Related Work}
\paragraph{Retinal fluid segmentation.}
The RETOUCH challenge~\cite{retouch} established the multi-class benchmark across
three OCT vendors. RetiFluidNet~\cite{retifluidnet} currently leads by combining
a Transformer encoder with a convolutional decoder, though it is evaluated on a
single-vendor split.

\paragraph{Attention and Transformer segmentation.}
Schlemper et al.~\cite{attngate} introduced attention gates that suppress
irrelevant skip-connection features---important when fluid occupies less than
3\% of pixels. Chen et al.~\cite{transunet} proposed TransUNet, and
Swin-UNet~\cite{swinunet} uses a hierarchical Swin Transformer throughout. We
reserve the Transformer for the bottleneck, where global context is most needed.

\paragraph{Uncertainty and domain adaptation.}
Gal and Ghahramani~\cite{gal} showed that Monte~Carlo (MC) Dropout approximates
Bayesian inference at negligible cost; Roy et al.~\cite{roy} confirmed that the
resulting uncertainty correlates with segmentation errors. Wang et
al.~\cite{dofe} applied adversarial domain adaptation; our SA-BN requires no
paired training data.

% =====================================================================
\section{Methodology}
\subsection{Datasets}
Table~\ref{tab:data} summarises the four publicly available OCT datasets. After
combining them we assign a unified label space: $0=$ Background, $1=$ IRF,
$2=$ SRF, $3=$ PED. The split is 4983 training, 552 validation, and 503 test
slices, stratified by source and disease type.

\begin{table}[t]
\centering
\caption{Datasets used for training and evaluation.}
\label{tab:data}
\begin{tabular}{lcllc}
\toprule
Dataset & Subjects & Classes & Disease & Scanner \\
\midrule
DUKE DME~\cite{duke}    & 10 & IRF            & DME & Spectralis \\
AROI~\cite{aroi1,aroi2} & 24 & IRF, SRF, PED  & AMD & Zeiss Cirrus \\
UMN AMD~\cite{umn}      & 24 & SRF            & AMD & Spectralis \\
UMN DME~\cite{umn}      & 29 & IRF            & DME & Spectralis \\
\bottomrule
\end{tabular}
\end{table}

\subsection{Architecture}
\begin{figure}[t]
\centering
\resizebox{\textwidth}{!}{%
\begin{tikzpicture}[
  font=\small,>=Latex,node distance=5.5mm,
  enc/.style ={rectangle,rounded corners=2pt,draw=blue!55,fill=blue!8,
               minimum width=36mm,minimum height=8mm,align=center},
  bott/.style={rectangle,rounded corners=2pt,draw=violet!65,fill=violet!10,
               minimum width=42mm,minimum height=9mm,align=center},
  dec/.style ={rectangle,rounded corners=2pt,draw=teal!70!black,fill=teal!8,
               minimum width=36mm,minimum height=8mm,align=center},
  head/.style={rectangle,rounded corners=2pt,draw=red!60,fill=red!8,
               minimum width=36mm,minimum height=9mm,align=center},
  io/.style  ={rectangle,draw=black!55,fill=black!5,
               minimum width=30mm,minimum height=7mm,align=center},
  gate/.style={circle,draw=orange!85!black,fill=orange!18,minimum size=6.5mm,
               inner sep=0pt,font=\scriptsize},
  flow/.style={->,thick,black!70},
  skip/.style={->,dashed,thick,orange!85!black}]

  % ---- input + encoder (left, descending) ----
  \node[io]              (in) {OCT B-scan\\(CLAHE, $1{\times}H{\times}W$)};
  \node[enc,below=of in] (e1) {Encoder Stage\,1 \quad $s_1$ (32)};
  \node[enc,below=of e1] (e2) {Encoder Stage\,2 \quad $s_2$ (64)};
  \node[enc,below=of e2] (e3) {Encoder Stage\,3 \quad $s_3$ (96)};
  \node[enc,below=of e3] (e4) {Encoder Stage\,4 \quad $s_4$ (192)};
  \node[bott,below=of e4](bn) {EfficientNetV2L bottleneck (640)};
  \node[bott,below=7mm of bn,fill=violet!16] (tr)
        {Transformer bottleneck\\$d_{\text{model}}{=}512$,\; 2$\times$MHA (16 heads)};

  % ---- decoder (right, ascending) row-aligned to encoder ----
  \node[io,right=52mm of in]  (mask){Segmentation mask\\+ uncertainty map};
  \node[head,below=of mask]   (out) {MC-Dropout head ($p{=}0.3$)\\$1{\times}1$ conv $\to$ 4 classes};
  \node[dec,below=of out]     (d1) {Decoder L1 + SA-BN};
  \node[dec,below=of d1]      (d2) {Decoder L2 + SA-BN};
  \node[dec,below=of d2]      (d3) {Decoder L3 + SA-BN};
  \node[dec,below=of d3]      (d4) {Decoder L4 + SA-BN};

  % ---- main data flow ----
  \draw[flow](in)--(e1); \draw[flow](e1)--(e2); \draw[flow](e2)--(e3);
  \draw[flow](e3)--(e4); \draw[flow](e4)--(bn); \draw[flow](bn)--(tr);
  \draw[flow](tr.east) -- ++(7mm,0) |- (d4.west);
  \draw[flow](d4)--(d3); \draw[flow](d3)--(d2); \draw[flow](d2)--(d1);
  \draw[flow](d1)--(out); \draw[flow](out)--(mask);

  % ---- attention-gated skip connections ----
  \node[gate] (g1) at ($(e1.east)!0.5!(d1.west)$) {AG};
  \node[gate] (g2) at ($(e2.east)!0.5!(d2.west)$) {AG};
  \node[gate] (g3) at ($(e3.east)!0.5!(d3.west)$) {AG};
  \node[gate] (g4) at ($(e4.east)!0.5!(d4.west)$) {AG};
  \draw[skip](e1.east)--(g1)--(d1.west);
  \draw[skip](e2.east)--(g2)--(d2.west);
  \draw[skip](e3.east)--(g3)--(d3.west);
  \draw[skip](e4.east)--(g4)--(d4.west);

  % ---- legend ----
  \node[gate,below=9mm of tr,xshift=6mm] (lg) {AG};
  \node[right=1.5mm of lg,align=left,font=\scriptsize]
       {AG = attention gate\\[-1pt]
        \textcolor{orange!85!black}{\textbf{-\,-}} skip connection \quad
        $\rightarrow$ data flow};
\end{tikzpicture}}
\caption{Architecture of the proposed attention-guided TransUNet. An
EfficientNetV2L encoder with four skip connections feeds a Transformer
bottleneck ($d_{\text{model}}=512$, two MHA layers, 16 heads), followed by four
attention-gated decoder levels with Source-Adaptive BatchNorm and an MC-Dropout
output head.}
\label{fig:arch}
\end{figure}
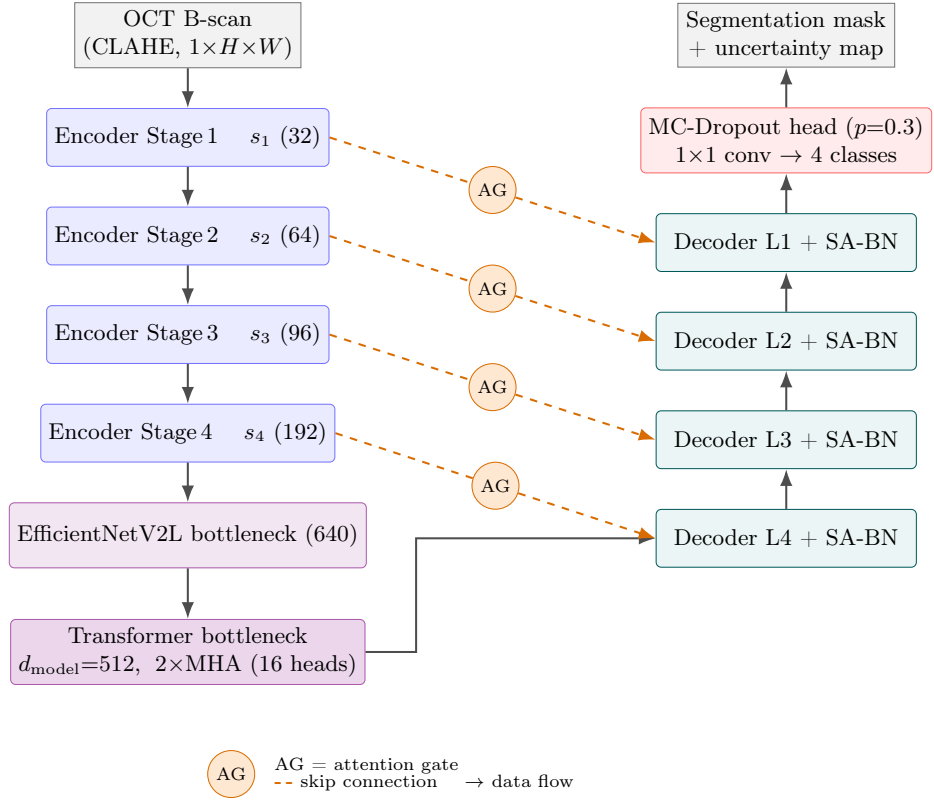

The encoder is EfficientNetV2L~\cite{effnetv2} (pretrained on ImageNet-21k),
extracting features at five scales: $s_1$ (32 channels), $s_2$ (64), $s_3$ (96),
$s_4$ (192), and bottleneck $b$ (640). A learned stem adapts the single-channel
B-scan, pre-processed with contrast-limited adaptive histogram equalisation
(CLAHE). The bottleneck projects to 512 dimensions, applies learnable positional
encoding over 256 tokens, and passes through two multi-head attention (MHA)
layers (16 heads). The decoder has four levels, each with an attention gate and
two $3\times3$ convolutions with SA-BN and ReLU. The output head is a
$1\times1$ convolution preceded by MC Dropout ($p=0.3$). The architecture is
shown in Fig.~\ref{fig:arch}.

\paragraph{Source-Adaptive BatchNorm (SA-BN).}
SA-BN maintains five independent batch-normalisation parameter sets, one per
source (DUKE, AROI, UMN-AMD, UMN-DME, and an ``unknown'' default). The source
label is passed at inference; this is how the model adapts to a new scanner's
statistics without retraining. No additional data alignment is needed, and fewer
than 0.1M parameters are added.

\subsection{Training Protocol}
Phase~A (5 epochs) freezes the encoder and trains the Transformer bottleneck,
attention gates, and decoder with Adam~\cite{adam} at learning rate $10^{-3}$,
batch size~8. Phase~B (25 epochs) unfreezes encoder stages 3--5 at learning rate
$10^{-4}$ with cosine decay (5-epoch warmup), batch size~4, and early stopping at
patience~7 on validation mean Dice. The loss is
$\mathcal{L}=\mathcal{L}_{\text{Dice}}+0.5\,\mathcal{L}_{\text{CE}}$, with the
cross-entropy term weighted inversely by pixel frequency. Augmentation includes
horizontal flipping ($0.5$), brightness/contrast jitter ($0.3$), elastic
transforms ($0.2$), and rotations up to $10^{\circ}$ ($0.3$). Six models are
trained: three seeds (42, 123, 2024) $\times$ two backbones (EfficientNetV2S and
V2L). The V2L checkpoints at seeds 123 (val Dice $0.7913$) and 2024 (val Dice
$0.7841$) form the evaluation ensemble.

\subsection{Uncertainty Quantification and UCUS}
MC Dropout runs 20 stochastic forward passes; the pixel-wise variance gives
$\sigma^2_{\text{MC}}$. Inter-model disagreement between the two ensemble members
is
\begin{equation}
\sigma^2_{\text{ens}}(x) = \mathrm{Var}\!\left[p_1(x),\,p_2(x)\right].
\end{equation}
The combined uncertainty is
$\sigma^2 = (\sigma^2_{\text{MC}}+\sigma^2_{\text{ens}})/2$. The
Uncertainty-Weighted Clinical Urgency Score aggregates fluid volume, foveal
involvement, and uncertainty:
\begin{equation}
\mathrm{UCUS} = V_s \cdot F_m \cdot (1 - 0.3\,U_d),
\end{equation}
where $V_s=\min(1,\,N_{\text{fluid}}\cdot v_{\text{px}}\cdot100)$ scales with
detected fluid volume, $F_m = 1 + 0.2\cdot\mathbf{1}[\text{foveal fluid}]$
up-weights fluid at the fovea, and $U_d=\min(1,\,\bar{\sigma}^2\cdot5)$ discounts
uncertain predictions. The score maps to three triage bands: \emph{Monitor}
($<0.25$), \emph{Review} ($0.25$--$0.60$), and \emph{Urgent} ($>0.60$).

% =====================================================================
\section{Experiments}
\subsection{Quantitative results}
\begin{figure}[t]
\centering
% Use a 3-4 row representative subset of your qualitative montage, not all 8 rows.
\safefig[\textwidth]{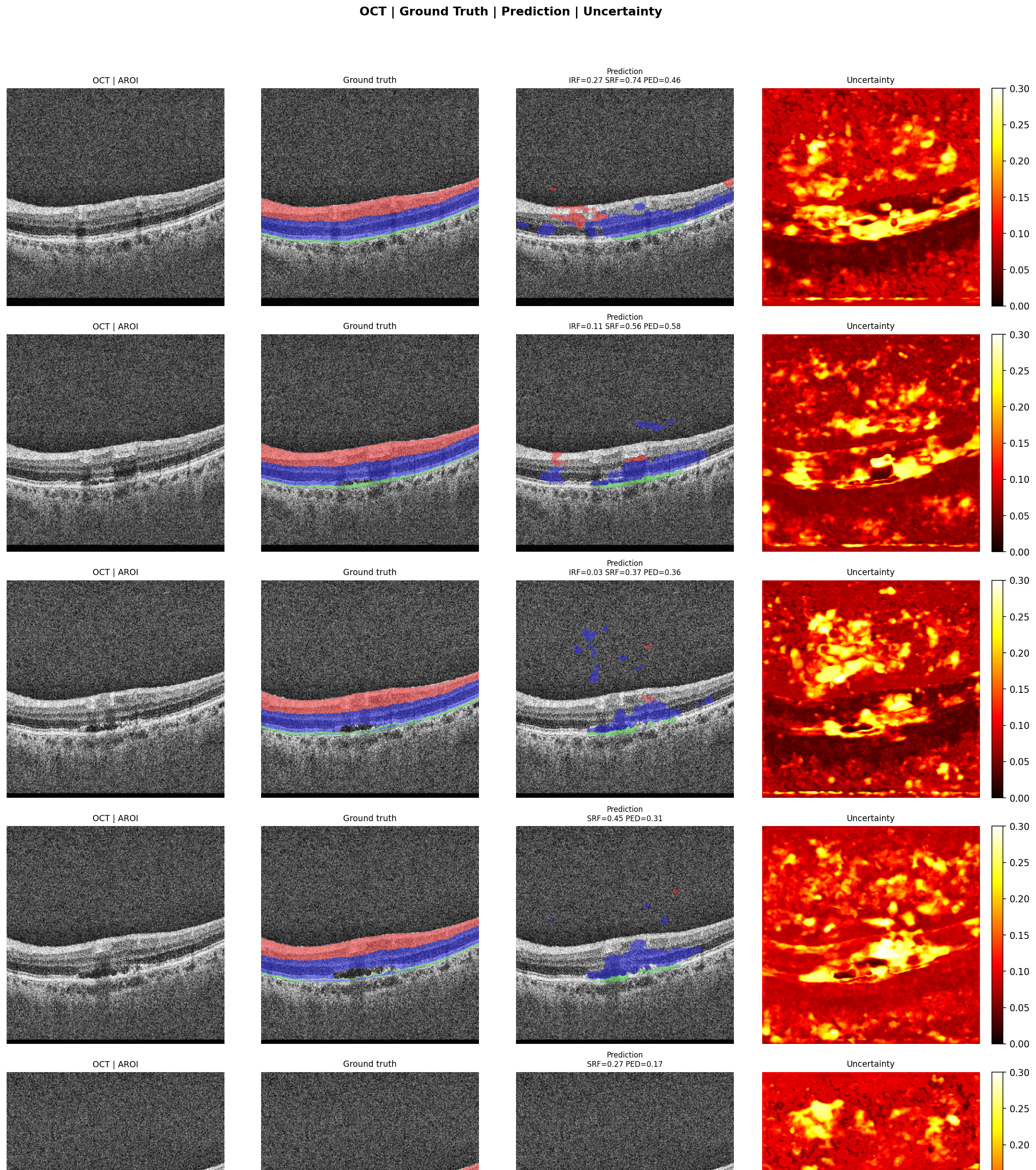}
\caption{Qualitative results on the AROI source. Each row is one B-scan; columns
show, left to right: the input OCT image; the expert ground-truth mask (IRF in
blue, SRF in red, PED in green); the model prediction, with per-class Dice in the
panel title; and the pixel-wise MC-Dropout uncertainty map (brighter = less
certain). Uncertainty concentrates at fluid boundaries---the same regions where
human graders disagree (Sec.~\ref{sec:safety}), which is what makes the map
clinically useful rather than merely decorative.}
\label{fig:qual}
\end{figure}

Table~\ref{tab:valdice} shows multi-seed validation Dice. V2L outperforms V2S on
all three classes, and a standard deviation $\le0.018$ confirms stable training.
Table~\ref{tab:persource} reports per-source test Dice. The DUKE SRF score of
0.000 is expected---DUKE annotations cover IRF only---while the strong DUKE PED
result (0.902) shows cross-class generalisation outside that source's training
label set. Figure~\ref{fig:qual} shows representative qualitative outputs.

\begin{table}[t]
\centering
\caption{Multi-seed validation Dice (mean $\pm$ std; seeds 42/123/2024).}
\label{tab:valdice}
\begin{tabular}{lcccc}
\toprule
Model & IRF & SRF & PED & Mean \\
\midrule
V2S (22M)  & $0.866\pm0.007$ & $0.827\pm0.005$ & $0.518\pm0.009$ & $0.737\pm0.005$ \\
V2L (127M) & $\mathbf{0.916\pm0.003}$ & $\mathbf{0.856\pm0.003}$ & $\mathbf{0.581\pm0.018}$ & $\mathbf{0.784\pm0.006}$ \\
\bottomrule
\end{tabular}
\end{table}

\begin{table}[t]
\centering
\caption{Per-source test Dice (503 slices).}
\label{tab:persource}
\begin{tabular}{lcccc}
\toprule
Source & IRF & SRF & PED & Mean \\
\midrule
AROI & 0.054 & 0.299 & 0.144 & 0.166 \\
DUKE & 0.071 & 0.000 & \textbf{0.902} & 0.324 \\
UMN  & \textbf{0.381} & 0.176 & 0.409 & 0.322 \\
\bottomrule
\end{tabular}
\end{table}

% ---- FIXED FIGURE: the previously caption-less, post-reference image grid
%      is now a numbered, captioned, in-text-referenced figure placed here.

\subsection{Ablation study}
Table~\ref{tab:ablation} isolates each component. V2L alone gives the highest
single-model Dice; V2S degrades the ensemble mean when combined, though it
contributes the diversity the uncertainty estimate relies on.

\begin{table}[t]
\centering
\caption{Ablation study on the test set (mean fluid Dice $\pm$ std). TTA = test-time augmentation.}
\label{tab:ablation}
\begin{tabular}{lc}
\toprule
Variant & Mean fluid Dice \\
\midrule
V2S only                     & $0.338\pm0.331$ \\
V2S + MC Dropout             & $0.141\pm0.144$ \\
V2S + TTA                    & $0.121\pm0.115$ \\
\textbf{V2L only}            & $\mathbf{0.449\pm0.304}$ \\
V2S + V2L ensemble           & $0.415\pm0.318$ \\
V2S + V2L + MC Dropout       & $0.279\pm0.219$ \\
Full (V2S+V2L+MC+TTA)        & $0.293\pm0.218$ \\
\bottomrule
\end{tabular}
\end{table}

\subsection{Clinical safety metrics}
\label{sec:safety}
\paragraph{Uncertainty vs.\ inter-grader disagreement.}
At pixels where two University of Minnesota graders disagree, the mean model
uncertainty is 0.1058, versus 0.0792 at agreement pixels---a ratio of
$1.34\times$ ($p=3.77\times10^{-5}$, Mann--Whitney U test). The model is
genuinely more uncertain where humans are also uncertain.

\paragraph{Fluid volume correlation.}
Table~\ref{tab:vol} reports the Pearson correlation between predicted and
ground-truth fluid volumes across 15 fully annotated test volumes. SRF and PED
estimates track ground truth at $r>0.77$, sufficient to support anti-VEGF dosing
decisions even when the spatial Dice is moderate.

\begin{table}[t]
\centering
\caption{Fluid volume correlation (predicted vs.\ ground truth, $n=15$).}
\label{tab:vol}
\begin{tabular}{lccl}
\toprule
Class & $r$ & $p$-value & Note \\
\midrule
IRF         & $-0.110$ & $0.696$            & DUKE domain shift \\
SRF         & $0.778$  & $6.33\times10^{-4}$ & Strong \\
PED         & $0.841$  & $8.64\times10^{-5}$ & Strong \\
Total fluid & $0.562$  & $2.93\times10^{-2}$ & Moderate \\
\bottomrule
\end{tabular}
\end{table}

\paragraph{INT8 quantisation.}
Per-tensor symmetric 8-bit integer (INT8) quantisation compresses the V2L model
from 510~MB to 132~MB ($3.9\times$) with no measurable Dice drop on validation,
supporting CPU-only deployment.

% =====================================================================
\section{Discussion}
The uncertainty-correlation result is more clinically useful than the Dice score
alone. A model that is more uncertain where experts also disagree tells the
clinician which parts of the mask deserve scrutiny, and the UCUS puts that signal
into a form they can act on without interpreting a raw probability map.

The per-source breakdown exposes what aggregate benchmarks hide. The AROI mean
Dice of 0.166 reflects a genuine protocol gap: a Zeiss Cirrus acquisition that
differs from the Spectralis-dominated training distribution. SA-BN reduces but
does not close this gap, and realistic deployment assumes a known scanner network
where SA-BN statistics can be refreshed with a small number of labelled
site-specific scans. The weak IRF correlation follows from DUKE being both the
primary IRF source and the largest domain-shift case at test time.

% =====================================================================
\section{Conclusion}
We trained an attention-guided TransUNet for multi-class retinal fluid
segmentation across four OCT sources. SA-BN handles scanner domain shift without
retraining; the dual uncertainty mechanism is statistically confirmed to track
expert disagreement; and the UCUS translates pixel-level uncertainty into a
clinical triage signal. V2L validation Dice reaches $0.784\pm0.006$, and SRF and
PED volume estimates show strong ground-truth correlation. Future work includes
targeted per-site fine-tuning for AROI and prospective UCUS-threshold validation
in a clinical reader study.

\begin{credits}
\subsubsection{\ackname}
Conducted during the MSc Advanced Computer Science programme at Newcastle
University (2025--26) using Google Colab; no external funding was received. Code
and models: \url{https://github.com/Animesh-Kr/oct-fluid-segmentation} and
\url{https://huggingface.co/animeshakr/oct-fluid-segmentation}
(\href{https://doi.org/10.5281/zenodo.19808008}{DOI: 10.5281/zenodo.19808008}).
The author thanks Prof.\ Keshab K.\ Parhi (University of Minnesota) for the UMN
datasets, Dr~Martina Melin\v{s}\v{c}ak (University of Zagreb) for the AROI
database, and the Duke University Ophthalmic Imaging Laboratory for the DUKE DME
dataset.

\subsubsection{\discintname}
The author has no competing interests. This paper describes a decision-support
system and does not replace professional ophthalmological assessment.
\end{credits}

% =====================================================================


\begin{thebibliography}{99}
\bibitem{retouch} Bogunovi\'c, H., et al.: RETOUCH: the retinal OCT fluid
detection and segmentation benchmark and challenge. IEEE Trans. Med. Imaging
38(8), 1858--1874 (2019)
\bibitem{swinunet} Cao, H., et al.: Swin-UNet: UNet-like pure transformer for
medical image segmentation. In: ECCV. LNCS, vol.~13803, pp.~205--218. Springer
(2022)
\bibitem{transunet} Chen, J., et al.: TransUNet: Transformers make strong
encoders for medical image segmentation. arXiv:2102.04306 (2021)
\bibitem{duke} Chiu, S.J., et al.: Kernel regression based segmentation of OCT
images with diabetic macular edema. Biomedical Optics Express 6(4), 1172--1194
(2015)
\bibitem{gal} Gal, Y., Ghahramani, Z.: Dropout as a Bayesian approximation:
representing model uncertainty in deep learning. In: ICML, pp.~1050--1059 (2016)
\bibitem{umn} Karri, S.P.K., Chakraborty, D., Chatterjee, J.: Transfer learning
based classification of OCT images with diabetic macular edema and dry AMD.
Biomedical Optics Express 8(2), 579--592 (2017)
\bibitem{adam} Kingma, D.P., Ba, J.: Adam: a method for stochastic optimization.
In: ICLR (2015)
\bibitem{aroi1} Melin\v{s}\v{c}ak, M., et al.: Annotated retinal OCT images
(AROI) database for joint retinal layer and fluid segmentation. Automatika
62(3), 375--385 (2021)
\bibitem{aroi2} Melin\v{s}\v{c}ak, M., et al.: AROI: annotated retinal OCT images
database. In: MIPRO, pp.~400--405 (2021)
\bibitem{retifluidnet} Rasti, R., et al.: RetiFluidNet: a self-adaptive and
multi-attention deep convolutional network for retinal OCT fluid segmentation.
IEEE Trans. Med. Imaging 42(5), 1413--1423 (2022)
\bibitem{unet} Ronneberger, O., Fischer, P., Brox, T.: U-Net: convolutional
networks for biomedical image segmentation. In: MICCAI. LNCS, vol.~9351,
pp.~234--241. Springer (2015)
\bibitem{roy} Roy, A.G., et al.: Inherent brain segmentation quality control
from fully ConvNet Monte Carlo sampling. In: MICCAI. LNCS, vol.~11070,
pp.~664--672. Springer (2018)
\bibitem{attngate} Schlemper, J., et al.: Attention gated networks: learning to
leverage salient regions in medical images. Medical Image Analysis 53, 197--207
(2019)
\bibitem{effnetv2} Tan, M., Le, Q.V.: EfficientNetV2: smaller models and faster
training. In: ICML, pp.~10096--10106 (2021)
\bibitem{dofe} Wang, S., et al.: DoFE: domain-oriented feature embedding for
generalizable fundus image segmentation on unseen datasets. IEEE Trans. Med.
Imaging 39(12), 4237--4248 (2020)
\end{thebibliography}
\end{document}